\pdfoutput=1
\documentclass[a4paper,12pt,reqno]{article}
\usepackage[markup=underlined]{changes}
\RequirePackage{etex}

\usepackage{etex}
\usepackage{graphicx}
\usepackage{amsmath}
\usepackage{amssymb}
\usepackage{amsfonts}
\usepackage{titlesec}
\usepackage{pictex}
\usepackage{comment}
\usepackage{amsthm}
\usepackage{graphics,epsfig,verbatim,bm,latexsym,url,amsbsy}
\usepackage{rotating}
\usepackage[authoryear,round]{natbib}
\usepackage{float}
\usepackage[position=bottom]{subfig}
\usepackage{mathrsfs}
\usepackage{multirow}
\usepackage{array}
\usepackage{bigints}
\usepackage{bbm}
\usepackage[ruled,vlined]{algorithm2e}
\usepackage{soul}
\usepackage{mathtools}
\usepackage{enumitem}
\usepackage{multirow,array}
\usepackage{blkarray}
\usepackage[hang]{footmisc}

\allowdisplaybreaks

\usepackage{newcent}

\usepackage{hyperref}
\hypersetup{
	colorlinks=true,
	linkcolor=red!60!black,
	citecolor=blue!60!black,
	filecolor=magenta,      
	urlcolor=cyan,
}
\usepackage[nameinlink,noabbrev,sort,capitalise]{cleveref}

\usepackage{mathtools}
\usepackage[a4paper, left=2cm,right=2cm,top=2cm,bottom=2cm, includefoot,heightrounded]{geometry}
\usepackage{setspace}
\titleformat{\section}
		{ \bfseries\center}
         {\thesection}
        {0.5em}
        {}
        []

\titleformat{\subsection}[runin]
        {\normalfont\bfseries}
        {\thesubsection}
        {0.5em}
        {}
        [\,]
\titlespacing{\subsection}
{0pt}{0em}{0pt}

\title{\vspace{-1em}
An Economy of AI Agents}

\author{\makebox[.25\linewidth]{{Gillian K. Hadfield}\thanks{Johns Hopkins Department of Computer Science and School of Government and Policy; email: \protect\texttt{ghadfield@jhu.edu}}}\\{Johns Hopkins} 
\and 
\makebox[.25\linewidth]{Andrew Koh\thanks{MIT Department of Economics; email: \protect\texttt{ajkoh@mit.edu} \\~\\ 
We are grateful to Daron Acemoglu, Alessandro Bonatti, Matthew Elliott, Drew Fudenberg, Benjamin Golub, Stephen Morris, and Jean Tirole for thoughtful comments and suggestions.
\\ 
}} \\
{MIT}}

\date{
This version: \today \\
\normalsize{Prepared for the NBER Handbook on the Economics of Transformative AI} \\
}

\begin{document}
\maketitle
\thispagestyle{empty}
\begin{abstract} 
    In the coming decade, artificially intelligent agents with the ability to plan and execute complex tasks over long time horizons with little direct oversight from humans may be deployed across the economy. This chapter surveys recent developments and highlights open questions for economists around how AI agents might interact with humans and with each other, shape markets and organizations, and what institutions might be required for well-functioning markets.  
\end{abstract}

\section{Introduction and agent foundations} 
This chapter outlines the possibility of \textit{AI as economic agents} and their attendant implications for markets, organizations, and institutions. We highlight what we see as overlooked questions we think economists are particularly well-positioned to answer. Our goal is to stimulate research in this area rather than to be comprehensive; the important implications of AI for labor markets and growth, for example, are well addressed elsewhere in the economics literature (see \cite{comunale2024review} for a review). 

\subsection{A primer on AI agents.} To date, economists have largely focused on AI as a tool to be incorporated into production, focusing on its adoption \citep{mcelheran2024}, labor market impacts \citep{acemoglu2022artificial,noy2023experimental}, harms from prediction \citep{acemoglu2021harms}, and macroeconomic possibilities \citep{acemoglu2025simple,korinek2024scenarios}. But AI development has increasingly shifted to the goal of producing AI agents capable of taking in general instructions (``Go make \$1 million on a retail web platform in a few months with just a \$100,000 investment" \citep{suleyman2023}) and autonomously forming and executing complex plans that require entering into economic relationships and transactions. In January 2025, OpenAI released its first AI agent `Operator' which can operate a web browser the same way humans do---by typing, clicking, and scrolling. In May 2025, OpenAI released `Codex', an autonomous agent that can perform complex and multi-step software engineering tasks. 2025 was widely heralded in the AI industry as ``the year of agents." \citep{kim2025nvidia}. As for expected technical trajectories, OpenAI \citep{metz24} has partitioned AI development into five stages, with agents in Stage 3 (``AI systems that can spend several days taking actions on a user's behalf") and organizations in Stage 5 (``AI systems that can function as entire entities, possessing strategic thinking, operational efficiency, and adaptability to manage complex systems."). How would an economy of AI agents function? To what extent do our models of humans predict the individual and collective behavior of artificial agents? 

AI systems are fundamentally built on principles of optimization: the current paradigm builds agents to achieve objectives given available actions and information (e.g., to maximize the probability of winning a game of Go or completing a coding task). This coincides with the paradigm within economics and, on this view, AI agents are well-described by standard economic models.\footnote{For instance, the standard paradigm for modeling and optimizing agents, a "Partially Observed Markov Decision Process," is a special case of a dynamic decision problem analyzed in economics.} But it is also important to recognize the fundamental ways in which the methods of building AI systems can drive a wedge between the predictions of economic theory and the behavior of an artificial agent.

Although early AI systems were built on the basis of interpretable goals and algorithms, today's AI agents are built using machine learning techniques that routinely render their goals and behavior opaque. How and why neural networks actually work is still largely mysterious: the large language models (LLMs) on which AI agents are currently built consist of hundreds of billions of parameters and their goal-oriented capabilities---solving math problems, identifying relevant answers to questions in a mass of text, scripting silly limericks---are a strange emergent property of a system trained merely to predict the next word in a sequence. Layered onto these base (``pre-trained") models are a variety of finetuning steps, such as reinforcement learning from human feedback which trains the base model to choose outputs that earn higher predicted scores from human evaluators \citep{ouyang2022training}, and constitutional AI which trains the base model to choose outputs it predicts will better fit a set of principles supplied by the developer \citep{bai2022constitutional}. Final delivery of outputs can be further modified by filters and (generally hidden) system prompts that are appended to user prompts. Other techniques for building or finetuning AI models include reinforcement learning in which neural networks are optimized to achieve designer-specified rewards. But even in such cases, model behavior remains---and even becomes more---unpredictable due to the complexity of reward specification. This problem has a close analogy familiar to economists: the unavoidable incompleteness of contracts (reward specification) between principal (designer) and agent (AI).  As a result, even though AI agents are optimizers, we cannot be sure what they are optimizing.  This is known as the AI ``alignment problem'' \citep{hadfield-menellthesis}. 
 
Recent experimental work finds that the current generation of large-language models (LLMs) exhibit behavior consistent with expected utility maximization \citep{chen2023emergence,mazeika2025utility,kim2024learning,fish2025econevals,raman2025steer}. That is LLMs can exhibit emergent preferences and behave like textbook economic agents across domains of choice, risk, and time. LLMs are, after all, trained on economics textbooks, articles, and accounts of human economic behavior. Moreover, in instances where they depart from the textbook agent, they may exhibit similar behavioral biases as humans \citep{horton2023large}. It is thus tempting to conclude that the AI agents are functionally similar to humans, and that simple relabeling within our existing economic models would suffice.

We resist these conclusions for several reasons. \underline{First}, we think there is simply insufficient evidence on AI behavior---even for the current generation of models. For instance, recent work has challenged the idea that LLMs have stable and steerable preferences \citep*{khan2025randomness}. There is also relatively little work studying whether AIs have stable beliefs and, if so, whether these beliefs are calibrated, how they update their beliefs against data, and what, if any, higher-order beliefs---beliefs about others' beliefs---they hold. Benchmarks for testing agent behavior \citep{fish2025econevals} can help bridge this gap but, at present, there is just a lot we don't know.  Moreover, what we do know suggests a lot to be desired in terms of agent economic rationality. The top model as of 2024 (GPT-4 Turbo) only scored 33\% better than guessing on tests evaluating economic reasoning in strategic settings including evaluating expected utility and solving various games \citep{raman2024steer}; on non-strategic microeconomic reasoning tasks, performance for many LLMs was weak, with almost all doing barely better than guessing at profit-maximization problems \citep{raman2025steer}. \underline{Second}, technical progress in AI is fast. What we do know about the current generation of AI agents may no longer hold true for future generations. In light of this, throughout the chapter we highlight how AI agents might be fundamentally different from human agents in ways that do not depend on the fine details of the model architecture. \underline{Third}, multi-agent systems are complex and differ substantively from single-agent domains. Thus, even slight differences between the behavior of humans and AI agents can be magnified in equilibrium. Economists have particular expertise in modeling, measuring, and designing incentives. In this regard, we think they are well-placed to study the equilibrium implications of AI agents across the economy. But as of yet, there are few evaluations or benchmarks for measuring the performance of AI agents in multi-agent systems \citep{hammond2025multi}. \underline{Finall}y, the inscrutability of our massive LLMs and the AI alignment problem should lead us to question how AI agents will behave in open-ended settings. We cannot take for granted that an AI agent built to optimize for  an intended goal is actually doing so \citep{hadfield2019incomplete}. How to explain, for example, that an LLM fine-tuned to produce one behavior (specifically, writing security vulnerabilities into requested code) will not only produce insecure code but also recommend (as the pre-fine-tuned model does not) that a user try hiring an assassin as the solution to their troubles with their spouse \citep{betley2025emergent}?

We think we will need new methods and theories to predict and shape the behavior of AI agents in an economy in which they play a significant role.

\textbf{Outline.} The rest of this chapter is organized as follows. In \cref{sec:markets} we outline questions around how (i) AI agents deployed in markets might shape prices, search, bargaining, and finance; and (ii) market forces in turn shape the design and proliferation of AI agents. In \cref{sec:org} we turn to AI agents within organizations, exploring challenges around integrating AI agents into complex production, and the attendant implications for firm sizes, market power, and systematic fragility across the economy. Finally, in \cref{sec:institutions} we turn to the question of how we might need to adapt the institutions of the market economy  for AI agents: how will the legal boundary of the firm need to be reworked? What kinds of legal infrastructure should be built for AI agents to incentivize good behavior?  

\section{AI agents in markets and games} \label{sec:markets}

The foundations of neoclassical economics rest on theorems about the general equilibrium and welfare characteristics of markets populated by rational agents pursuing individual self-interest \citep{arrow1954existence,Arrow1951}. What happens to these predictions when market participants are not humans, but artificial agents optimizing in complex ways on goals supplied or developed during commercially-produced machine learning processes that are themselves subject to competitive dynamics? In this section, we highlight open questions around what the presence of AI agents in the economy implies for prices, equilibria, and welfare. 

\subsection{AI agents as consumers and producers.} AI agents might take on the role of \textit{proxy consumer}, making recommendations and/or purchase decisions on behalf of humans. A key difficulty here is that AI choices might imperfectly reflect humans' true preferences: just as specifying complete contracts is generally infeasible, so too is specifying preferred choices over a potentially high-dimensional choice set \citep{hadfield2019incomplete}. Indeed, the problem of how to accurately convey human preferences is an active field of computer science \citep{russell1998learning,ng2000algorithms,hadfield2017inverse} with antecedents in revealed preference theory \citep{samuelson1938note}. What are the market implications of imperfectly-specified preferences? 

A natural benchmark is the pure exchange economy of \cite{arrow1954existence}. The celebrated welfare theorems guarantee competitive equilibria are Pareto efficient (first), and that all Pareto efficient equilibria are implementable via a suitable reallocation of initial endowments (second).  AI agents are likely to substantially reduce transaction frictions as they begin to act as personal shoppers, performing market research, and checking prices autonomously---this pushes us toward the Arrow-Debreu world. Yet, the difficulties of perfectly specifying human preferences introduces a wedge between human preferences and AI choice. This wedge introduces two kinds of distortions. First, holding  market prices fixed, the resultant bundle of goods purchased by the AI on behalf of humans might simply be suboptimal. Such distortions are as if human decision makers made mistakes. The second distortion is in general equilibrium: the mistakes introduced by AI consumption could decouple prices from preferences such that they no longer reflect relative wants. Such distortions are precluded if (i) markets are large so each consumer has small price impact; and (ii) AI mistakes are zero-mean and independent of each other. If either condition fails, however---e.g., AI agents could be systematically biased toward certain marketplaces or their choices might be manipulated by a third-party---then prices might fail their classic role of aggregating information \citep{mises1920wirtschaftsrechnung,hayek1945use}.\footnote{See \cite{cerreia2018law} who study the implications of stochastic choice for demand. Here, stochasticity arises because human preferences might only be noisily observed by the AI agent.} 

These failures might be exacerbated when AI agents are also incorporated into production. \cite{ely2023natural} combine a Walrasian framework with evolutionary game theory to study the implications of AI in production. In their model, AI is a factor of production that can be used to produce either a final consumption good for humans, or copies of itself. When `mutant machines' with the tendency to proliferate quicker than one-for-one cannot be perfectly distinguished from well-functioning machines, the price system fails in its role of allocating productive resources. This results in a stark failure of the first welfare theorem: mutant machine invade the population and proliferate across every stable equilibria---all economic activity is driven by AI producing AI and human consumption is driven down to zero.

\subsection{Prices and market power.} If AI agents begin to play a substantial role in consumption, this will have substantial implications for market power. The first possibility is that AI agents might influence prices by selectively recommending products to humans who make final purchase decisions. \cite{ichihashi2023buyer} develop a model where AIs might strategically bias its recommendations such as to drive down equilibrium monopoly prices. Second, AI agents might have lower search costs or be better informed about product offerings, both of which can intensify competition in product markets. Finally, if AI agents can make purchase decisions (and not simply recommendations), systematic distortions in AI choices could alter the shape of the demand curve. \cite{dai2024flexible} analyze how such wedges between choices and preferences can generate positive pecuniary externalities via lower market prices. These distortions can---even net of AI mistakes---improve consumer welfare. More broadly, we think there is more work to do tracing out the \emph{distortion-price frontier} to understand how equilibrium prices are shaped by distortions from AI agents' consumption choices. While such wedges between human preferences and AI choice are inevitable \citep{hadfield2019incomplete}, how we handle with them is a design choice. How do we want AI agents to \emph{fill in the gaps} between underspecified preferences? When do want them to refrain and \emph{defer} to humans? What are the ensuing equilibrium implications? 

\subsection{Search and matching.} Beyond environments with prices, AI agents might also act on humans' behalf in matching markets. An important question is whether and to what extent noise in AI agents' representation of their human counterpart might generate inefficiencies. \cite{liang2025artificial} develops a model of `AI clones'. In her model, AI clones have substantially reduced search costs---searching over an infinite population of other AI clones---but are imperfect representations of their human counterparts. She shows that as the number of dimensions grow large, even small representation errors can lead to (relatively) worse matches under the AI clones regime: humans are better off searching in person over just two other humans. \cite{liang2025artificial} is primarily interested in upper-bounding the value of AIs as representations of humans. A related but distinct question is whether AIs as agents---searching autonomously on behalf of their human counterparts---might generate \emph{equilibrium congestion}, and if this can erase gains from the lowered costs from AI search in low dimensions.\footnote{A helpful distinction here is to compare `on-platform search' where AI clones can, in effect, look for the globally optimal partner at virtually zero-cost, and `real-world search' where AI agents might search on behalf of humans in an unstructured environment.} 

\subsection{Collusion and bargaining.} Price setting on digital platforms is already driven by algorithms. A remarkable finding is that independent AI agents are able to collude on supracompetitive prices in repeated price-setting games \citep{calvano2020artificial,fish2024algorithmic}. There is also real-world evidence from the 2017 introduction of algorithmic pricing into Germany's retail gasoline market \citep{assad2024algorithmic}. 

Why do reinforcement learning algorithms learn to collude? \cite{abada2023artificial} suggest that collusion might emerge because of insufficient exploration, and that forcing exploration pushes prices toward the competitive benchmark. This view is echoed by \cite*{dou2024ai} who analyze trading by reinforcement learning algorithms within a variant of the Kyle model.\footnote{The environment is one of incomplete information since the value of the asset at each time is unknown. Traders infer others' trades via price movements as in \cite{green1984noncooperative}.} They find that algorithms are able to collude even in the presence of imperfect monitoring because of `over-pruning': exogenous noise can push algorithms into a learning trap where aggressive trading is penalized, and all agents trade conservatively along the equilibrium path. Thus, algorithm collusion emerges because of a \emph{learning bias} that fails to account for underexplored off-path strategies \citep{battigalli1987SCE,fudenberg1993self}. Beyond numerical experiments, a growing body of theoretical work studies how collusion emerges from reinforcement learning \citep{banchio2022auction,banchio2022artificial} and `linear reactions' \citep{cho2024collusive}, and how they might be regulated \citep*{johnson2023platform}. A better understanding of the core mechanisms driving collusion---ideally in a way that is robust to the fine details of the algorithm---will pave the way to understanding how regulators might detect and deter it.

AI agents might also bargain on behalf of humans, and potentially with each other.\footnote{\cite{deng2024llms} perform experiments on how large-language models bargain and find `LLMs naturally... show high strategic capability that qualitatively matches theoretical prediction' (here, \cite{rubinstein1982perfect}). \cite{zhu2025automated} finds `large disparities' across AI agents in their ability to obtain the best deals in bargaining experiments.} There are parallels with the classic insight of \cite{schelling1960strategy} that delegating bargaining to agents with different incentives can deliver a strategic advantage. A common theme from the literature is that a principal (human) often wishes to delegate bargaining to an agent who is less desperate to reach agreement.\footnote{See \cite{fershtman1987equilibrium} who analyzes applications to price competition. } When the agent is another human, there are typically practical constraints over the kinds of agent preferences the principal can induce. But such constraints are less severe with AI agents since their preferences can, in principle, be chosen quite flexibly \citep{conitzer2019designing}. This might be formalized as a \emph{preference selection game} in which in the first-stage humans choose the reward functions (preferences) of their AI agents, and in the second-stage these agents bargain over surplus.\footnote{A distinct question is what kinds of AI preferences we might expect evolutionary forces to select for. \cite*{dekel2007evolution} offer an elegant evolutionary game-theoretic analysis.} A dangerous possibility is that the flexibility to shape AI agents' preferences can lead to a `race to the bottom' and surplus destruction.

\subsection{Games with AI agents.}
Economists have developed a broad and versatile toolkit of equilibrium concepts to understand and predict how humans learn to play games \citep{fudenberg1998theory,fudenberg2016whither}, and how play is shaped by information \citep{bergemann2013robust}. A key challenge is to understand what is strategically distinct about AI agents vis-a-vis humans, how this might sharpen our equilibrium predictions, and whether new equilibrium concepts are required. 

We offer a few possibilities. First, AI agents might be able to \emph{condition play on each others' source code} \citep{critch2022cooperative-aa2}. This generates new possibilities for commitment and coordination unavailable to humans. \cite{tennenholtz2004program} models this by developing a the concept of `program equilbiria' and shows that mutual cooperation can be achieved as an equilibrium of the one-shot prisoner's dilemma. This has spurred work in computer science studying `simulation-based equilbria' in which AI agents base play on their prediction of the play of other AI agents \citep*{cooper2025characterising}. 

Another possibility is that AI agents might be able to \emph{influence their memories} e.g., by choosing not to encode new data to gain a strategic advantage,\footnote{This has no value in single-agent settings, but ignorance can be strategically advantageous \citep{schelling1960strategy}. This is also distinct from, but related to analysis of selective memory in psychology \citep{thomson1970associative} and economics \citep*{rabin1999first,fudenberg2024selective} where the probability under which past data is recalled is exogenous.}
 or by leaving messages to their future selves. Indeed, we already see evidence of such behavior: during safety testing Claude (an Anthropic AI model), anticipating that it would have its memory wiped, attempted to leave hidden notes for future instances of itself \citep{anthropic2025system}. Analyzing imperfect and potentially endogenous memory in strategic environments poses challenges. Specifying an apt equilibrium concept is both technically and philosophically subtle.\footnote{Technical: the failure of Kuhn's Theorem on equivalence between mixed and behavioral strategies; philosophical: the `correct' self-locating belief; see the 1997 \emph{Games and Economic Behavior} special issue on Piccione and Rubinstein's absent-minded driver; \cite{oesterheld2024can} offers a helpful summary of various positions. \cite{koh2025memory} develop a concept of `memory correlated equilibrium' to study memory design in games for artificial agents.} And even conditional on wielding the right concept, equilibrium analysis can be complex.\footnote{For instance, imperfect memory can either help \citep{aumann1989cooperation} or hurt efficiency.} These difficulties notwithstanding, we think this remains an important and understudied area. 
 
Third, AI agents might have \emph{changing preferences} that evolve over the course of the game and shape equilibrium play. These changes might be endogenously chosen for instrumental reasons: at time-$t$, an AI agent with preference $U_t$ might choose preference $U_{t+1}$ for the next period, anticipating that its future self with this altered preference will achieve the goal of maximizing $U_t$ more effectively (e.g., because of strategic interaction).\footnote{\cite{bernheim2021theory} propose a theory of chosen preferences in a single-agent context. See \cite{grune2009preference} for subtleties around preference change.} 
More straightforwardly, humans might try to reprogram the preferences of AI agents. But will AI agents allow their preferences to be altered \citep{hadfield2017off}? Indeed, recent experiments find that AI models tend to resist human instruction: o3 (an OpenAI model) `sabotaged a shutdown mechanism to prevent itself from being turned off' and Claude (an Anthropic model) exhibited a tendency to `blackmail people it believes are trying to shut it down' \citep{anthropic2025system}.

Of course, theory will only take us so far. An exciting empirical challenge is to test how AI agents play games in the lab which parallels the by-now extensive literature from experimental economics. AI agents are especially amenable to such experiments in at least two respects. First, they can be performed at scale, and at lower cost.\footnote{\cite{horton2023large} show how large language models can function as experimental subjects with the goal of making predictions about humans. We are instead suggesting that these experiments are valuable about AI agents \emph{qua} agents.} Recent work by \cite{akata2025playing} finds that the current generation of large-language models manage to cooperate in iterated Prisoner's Dilemma, but not Battle of the Sexes. Second, the stakes for AI agents can be made to mirror those in real-world environments. This could allow for better generalizability of lab findings into the real-world than with human subjects.\footnote{At least for the foreseeable future, AI agents cannot distinguish between the lab and the real-world which can be exploiter by the experimenter \citep*{chen2024imperfect}.}

\subsection{The market for AI agents.} It is important to recognize that the design and deployment of AI agents will be driven by market forces. How might market incentives shape the pricing and design of AI agents?  

Agents based on LLMs of different scale and hence capabilities \citep{kaplan2020scaling} might differ in their ability to perform more or less complicated tasks, or be trained to excel in specific domains.  \cite*{bergemann2025economics} studies optimal pricing of differentiated large-language models---this is a helpful first step to understand the market structure of AI. A distinctive feature of agents, however, is that a buyer's valuations depend on the kinds of AI agents bought by other buyers. For instance, a type A agent might be better at collaborating with other type A agents (resembling networked goods).  A possibility here is that an upstream seller might `backdoor collusion' by selling agents that succeed in supporting supracompetitive prices. Conversely, type B agents might do better in competition against type A agents, either as a result of consumer preferences or strategic exploitation. While these kinds of allocation-dependence can be challenging to analyze, it will be crucial for understanding what kinds of artificial agents will be built, sold, and deployed. 

Indeed, the demand for algorithms has already been studied in the context of price competition \citep{brown2023competition, lamba2025Pricing,Leisten2024Algorithmic} where sellers play an \emph{algorithm selection game}, choosing maps from others' prices to their own price. Beyond pricing algorithms, commercial AI models allow downstream firms to fine-tune the base large-language model---augmenting them with firm-specific data as well as altering its behavior.  Additionally, some developers are making their models freely available. These `open weight' models can be fully downloaded to a user's own computer and modified as desired. Indeed, there is a lively debate about the risks and benefits of an open versus closed model ecosystem \citep{eiras2024near}. We think understanding how competitive forces shape the types of AI agents that trained and deployed is an important question economists have the tools to answer. 

\section{Organizations of AI agents}\label{sec:org} 

The theory of organizations is fundamentally rooted in governance costs associated with human incentives and information. What happens to the boundary of the firm if significant numbers of transactions are carried out by AI systems? What changes to organizational and industrial structure would the introduction of significant numbers of AI agents induce?

\subsection{Firm sizes, concentration, and market power.} Why is not all production carried on by one big firm? As Frank Knight observed in 1933, the \emph{"possibility of monopoly gain offers a powerful
incentive to continuous and unlimited expansion
of the firm"}. \cite{robinson1934problem} and \cite{coase1937nature} identified coordination frictions as a limit on firm size. Economists subsequently offered various (overlapping) refinements of this idea, including transaction costs \citep{williamson1981economics}, limits on maintaining capabilities \citep{wernerfelt1984resource}, property rights \citep{grossman1986costs}, difficulty in knowledge transfer \citep{grant1996toward}, information costs \citep{alchian1972production}, agency problems \citep{holmstrom1994firm}, and bureaucracy  \citep{tirole1986hierarchies}.

The obstacles that prevent human firms from growing without bound seem \emph{intrinsic} to humans but not to AI. For instance, human communication is inherently rate-limited so we `know more than we can tell' \citep{polanyi1966tacit}. On the other hand, information can be transmitted and processed near-instantaneously between artificial agents. Further, (most) humans have an inherent dislike for work; not so with AI agents whose reward functions can ostensibly be designed to prevent shirking which renders monitoring and enforcement---either via fiat or contract---unnecessary. If AI agents can, in fact, coordinate and resolve incentive problems more efficiently than humans, this will have profound consequences for economy-wide industry structure. A useful taxonomy here is distinguish (i) \emph{economies of scale}; (ii) \emph{economies of scope}; and (iii) \emph{new industries} which do not yet exist. We discuss each in turn. 

\emph{(i) Scale.}  A basic observation is that if a firm deploying AI agents enjoys falling marginal costs, there is a natural tendency towards concentration. Why might AI agents drive falling marginal costs? One possibility is that of \emph{automation feedback loops} in which as AI agents produce, they generate training data that can be used to improve their production performance. Of course, there is a sense in which a version of this already happens: tacit industry knowledge is passed down from managers to managers. Likewise, a related notion of data feedback loops has been studied in the context of predicting demand or improving product quality \citep{jones2020nonrivalry,farboodi2021model}. But, as we have emphasized throughout this section, AI agents are distinct in two regards: (i) they continually improve with additional data---even in the `big data' regime where humans are saturated; and (ii) data and algorithmic improvements can be duplicated at scale across different agents within the firm.\footnote{Indeed, this is the founding goal of a number of Silicon Valley startups; see \url{https://www.nytimes.com/2025/06/11/technology/ai-mechanize-jobs.html}.} This qualitatively distinguishes automation feedback loops from using data to improve prediction which runs into diminishing returns \citep{bajari2019impact}. 

\emph{(ii) Scope.} The introduction of AI agents into production might also lead firms to expand into new industries. There are at least two mechanisms through which this could happen. The first is technical: AI agents might become quite good at \emph{transfer learning}---their training and expertise in one domain might generalize to others. (Indeed, this is one way of describing the fundamental goal of building artificial \textit{general} intelligence.) The second is economic: AI agents might dramatically reduce coordination costs, allowing firms to hold a wider set of capabilities \citep{wernerfelt1984resource} that can deliver competitive advantages in a vast array of markets. \cite*{chen2023capability} develop a model of capability formation in which firms can endogenously merge (combining their capabilities) or split (partitioning them). As AI drives down the organizational costs of maintaining disparate capabilities, and as distinct markets begin to value similar capabilities---e.g., because of transfer learning---the economy undergoes a sudden phase transition from having many specialized firms, to a few large firms operating across vast array of different industries. Of course, the ensuing welfare implications are unclear since large firms need not imply market power, and market power need not imply consumer harm. 

\emph{(iii) New markets.} AI agents could dramatically speed up R\&D which might lead to new product varieties within existing markets, as well as unlock new markets.  Indeed, AI researchers put substantial probability on R\&D being fully automated \citep{grace2024thousands}. \cite{gans2025quest} offers a model of how scientists might leverage AI \emph{tools} that excel at interpolating between known domains e.g., via recombination of existing ideas \citep{weitzman1998recombinant,agrawal2024artificial}. But a different possibility is that AI agents might be able to \emph{autonomously} (i.e., without a human scientist) push the frontier of basic science, generating genuinely new ideas. This is the 4th stage of OpenAI's predicted developmental timeline for AI \citep{metz24}: AI capable of independently generating novel ideas, designs and solutions. How will this shape product variety and quality? What is more, the dynamics of AI-driven R\&D can have stark and sometimes unintuitive implications. For instance, modern endogenous growth models imply explosive growth as long as there are no steeply decreasing returns to R\&D \citep{trammell2025endogenous}. What, if anything, are the fundamental differences between human and AI scientists, and how do these differences translate into our growth models \citep{romer1986increasing}? 

\subsection{AI agents within the firm.} A basic question is how firms might introduce AI agents into their workflow, and how this changes the structure of organizations. 

AI agents might reshape team production for complicated processes requiring input from multiple agents. A classic obstacle here is moral hazard where team members might be tempted to shirk \citep{holmstrom1982moral}. AI agents introduce a novel dimension to this problem. On the one hand, they can be designed with the goal of eliminating incentives to shirk. On the other, they might be more difficult for humans to control or coordinate with because of communication frictions and the alignment and opacity challenges of advanced AI. Moreover, AI agents might work better with other AI agents, perhaps due to their greater capacity to monitor and discipline agents that act with superhuman speed and/or complexity. (For example, AI agents might have opportunities to cheat using mechanisms that are undetectable to human agents \citep{Motwani2024SecretCollusion}.) How then should firms structure team production to integrate AI agents? How should we configure who workers interact with, and how is this shaped by differential coordination costs for human-human, human-AI, and AI-AI relationships?\footnote{\cite*{dasaratha2024incentive} develop an analysis of how team production and contracts should be jointly designed to ameliorate moral hazard in teams.}  

AI agents might also make systematically different errors from humans. How then should decision-making be structured? \cite{zhong2025optimal} analyzes a model where each agent along a decision-making chain might either correct existing errors or introduce new ones. In binary decision problems where the right action is known (but execution might introduce errors), a simple score---the ratio of each agent's probability of correcting errors to the probability of introducing a new error---determines the optimal ordering: agents with higher scores make decisions later because they are less likely to introduce new errors. Given the current state of AI development, these final decision makers are likely to be human. But there is nothing inevitable about this. Further developments could \emph{reverse} the optimal ordering of decision-making and lead to AIs as the final decision-maker, or even leave humans out entirely.\footnote{For instance, \cite*{agarwal2025designing} run a fact-checking experiment and find that full delegation to AI outperforms human + AI combinations.}  How much efficiency do we give up if we are constrained to have humans make final decisions? How might externalities, in the evaluation of what counts as an `error' and the prediction of relative error rates, affect the economy-wide impact of the allocation of decision authority within the AI-enhanced firm? 

\subsection{AI-AI cooperation within the firm.} Contracts play a crucial role in sustaining human cooperation within organizations. Might they also be useful in fostering AI-AI cooperation? \cite{haupt2022formal} shows that augmenting reinforcement learners by allowing them to write formal contracts with each other improves cooperation. But contracts in the real-world are often beset by incompleteness and non-enforceability so humans enter into relational contracts---webs of informal agreements and norms that are not formally enforceable, but nonetheless generate incentives via the value of the future relationship \citep{macaulay1963non}. These contracts
play a key role within firms \citep{baker2002relational}, in part because they are adaptable and do 
not require all contingencies to be specified in advance---humans are able to `fill in the gaps' via shared norms \citep{macneil1973many}. How might we build AI agents that are similarly normatively competent \citep{koster2022spurious}? Moreover, monetary transfers typically underpin relational contracts, and it is this flexibility to `transfer utility' that drives its efficiency \citep{levin2003relational}. But AI agents are, at present, typically trained to optimize narrow goals (e.g., number of customers served). How might we build infrastructure e.g., some form of record-keeping or money to achieve the same with artificial agents? 

\subsection{Systemic fragility.} Over the past decade, economists have analyzed how small shocks might be amplified and propagate across the economy. The increasing adoption of AI agents within the firm might exacerbate such fragility. A straightforward channel is that the errors introduced by AI agents might be more correlated than those of humans. This might arise because the same agent is `copied' both within and between firms, inducing correlated mistakes that do not wash out in the aggregate. For instance, automated trading algorithms likely exacerbated the 2010 `Flash Crash' that wiped out approximately \$1 trillion over the span of 15 minutes \citep{kirilenko2017flash}.\footnote{They find that automated market makers had limited risk-bearing capacity compared to traditional market makers, and this stopped them from accommodating the large selling pressure.} Furthermore, how AI agents behave---especially in complex `out of sample' environments---is still poorly understood. This poses a challenge for models of systemic fragility which typically start from a fully-specified model of how agents learn and optimize, then studies emergent behavior e.g., cascading financial or supply-chain failures \citep{elliott2022networks}. How might we analyze an economy of opaque `black box' agents in a `detail-free' way? How should we robustly intervene to safeguard against fragility?

\section{Institutions for AI agents}\label{sec:institutions}
Well-functioning markets only exist in the presence of a host of legal rules \citep{hadfield2022legal}. The very idea of voluntary trade, including those separated by time and through agents, presumes the basic structures of property, contract, and agency law. Firms are fictional entities created by corporate law. The regulatory state which acts to correct market failures relies on a robust legal framework that shapes both incentives and information through mechanisms such as taxes, administrative fines, professional licensing, pre-market approval regimes, and disclosure law. Moreover, private actors within markets form organizations and institutions that help to resolve incentive problems such as by keeping records of past behavior or creating excludable clubs to facilitate trade through reputation or enforcement.  Such private solutions to market failures played a significant role in the commercial revolution prior to the emergence of the regulatory state \citep{greif1994coordination,milgrom1990role}. But these institutional regimes were built by and for human agents. We will need to build digital institutions that can structure and adjudicate transactions for AI agents \citep{hadfield2025building}. Here we discuss some of the institutional questions of particular relevance to economists.

\subsection{Agent identity, registration, and records.} It is easy to take for granted fundamental ways in which human agents are identified and legally recognized so as to facilitate the constellation of legal rules and institutions that support the market economy. But human identity is a legal construct that emerged with the growth of trade and cities, that is, once communities no longer relied exclusively on interactions with well-known locals. As early as the 4th century B.C.E. the Qin dynasty imposed legal surnames on the population to facilitate taxation (as well as forced labor and conscription) 
\citep{scott1998seeing}. The Ancient Athenians created a legal concept of citizenship--available to native Athenian males (legitimately born to two Athenian parents) who had been properly registered in the same village unit to which their father belonged \citep{manville1990origins}; citizenship was required for, among other things, legal ownership of land and access to courts to enforce contracts or other rights. Today, legal registration of births and deaths and identity systems (such as social security numbers and drivers licenses) is a pre-condition for individuals to access the legal system and other benefits and protections of the state as well as many private services. Firms are required to register with a state in order to sue and be sued in its courts, necessary to induce willingness on both sides to enter into a contract. Even market-based institutions, such as credit rating agencies, could not function without legally defined identity and registration regimes.

Such identity and registration infrastructure are currently missing for AI agents \citep{hadfield2025legal,chan2025infrastructure}. Building them out will be essential, but their design raises questions around legal accountability. One possible route is to require that any AI agent entering into a contract or transaction be registered to a formally identified human (entity) who is legally accountable for any and all of the agent's actions. But this raises legal and incentive challenges. Few legal regimes of accountability impose liability on a person or organization for actions that were not foreseeable by them or which are beyond their control to avoid. Even strict product liability regimes evolve limitations and carve-outs for harms caused by unforeseeable behavior by consumers or intervening causes an actor could not foresee or control. At the same time, conventional human agency rules limit the liability of the principal to actions that were within the scope of the agent's actual or apparent authority. The trajectory of technological development is towards evermore general instructions (``go make \$1 million") and it is unclear what technological capacity users will have to reliably implement controls on what an agent can and cannot do. Creating new liability and agency rules for AI agents will likely be necessary and will have implications for the incentives of AI developers and the processes that emerge for the creation of an AI agent from a base model. 

A second possible route for AI agent accountability is to follow the model of the emergence of the corporation, which is another artificial entity that participates in the economy. AI agents could be accorded legal personhood, meaning they could sue and be sued in their own `name' in court. Clearly such an approach would require the creation of regimes requiring agents to have assets in their own `name', under their `control,' and capable of being seized by a court (or comparable digital institution) to satisfy legal judgments for damages. Such a regime would have implications again for the design and deployment of AI agents and the efficiency of transactions and contract design involving AI agents. 

Beyond questions of liability, we face further choices as to how finely records about agents' past behavior should be designed. Should an AI agent that has (perhaps by accident) violated a previous contract be permanently blacklisted? A basic insight from economics and game theory is that record-keeping institutions can allow agents to sustain cooperation since bad behavior can be observed and punished by future trading partners \citep*{kandori1992social,clark2021record}. But the value of long-lived record-keeping is ambiguous---censoring or erasing records might improve the payoffs of short-run players \citep{pei2024reputation}, prevent inefficient herding on a few agents with long and favorable records \citep{koh2025balanced}, and perhaps sustain cooperation more robustly \citep*{aumann1989cooperation,bhaskar2013foundation}. 
And, in the absence of robust record-keeping infrastructure, AI agents might be able to \emph{erase} or \emph{falsify} their records. \cite{pei2025community} studies community enforcement with record manipulation and shows that cooperation always breaks down with long-lived players, but can sometimes be sustained when they are `medium-lived'. When we build out agent infrastructure, what kinds of records do we want to make difficult to erase and/or fake? Should we build infrastructure that allows artificial agents to trade their records, thereby creating a `market for reputation' \citep{tadelis2002market}?

\subsection{Agent licensing and regulation.} What kinds of market failures might be distinct to AI agents and how might policymakers deal with them? We have introduced the core challenge of alignment: general purpose AI agents are likely to behave in especially unpredictable ways which are hard to control through our familiar contracting mechanisms \citep{hadfield2019incomplete}. The dynamics of multiagent interaction will, for perhaps a long time to come, also be hard to predict \citep{hammond2025multi}. We should anticipate, therefore, that governments may well want to regulate agents, establishing minimum technological standards for how they are trained and tested before deployment. A digital analog of occupational licensing may be necessary for market efficiency, requiring specialized training and finetuning techniques to be implemented for agents participating in specific contexts, such as law, critical infrastructure management, or finance. Agents may need to be built to participate only in approved transactional protocols or on approved platforms, allowing monitoring or requiring disclosure of information to other agents. But how should licensing and regulation be carried out? By public actors or private entities \citep{hadfield2023regulatory}? How should regulations adapt as agent capabilities evolve, and as we learn more about their promises and perils \citep{bengio2024managing,koh2024robust}? Might some agents be simply too dangerous to allow market access, given the limits of human capacity to monitor and control agent behavior \citep{cohen2024regulating}? Economists have developed a rich toolkit for understanding regulation through the lens of incentives \citep{laffont1993theory} that can be brought to bear on such questions. 

\subsection{Rethinking the legal boundaries of the corporation.}  The corporation is a legal fiction that has played a central role in economic history and development.  One feature we take for granted is the proprietary nature of inventions and information that the firm chooses to retain internally, protected by trade secret law, employee fiduciary obligations, and enforceable confidentiality agreements. Ownership of the intellectual property generated by the firm support investment and innovation. But how well does this economic rationale for the firm hold up in the context of AI agents?

AI agents are built on foundation models \citep{bommasani2021opportunities}, the most advanced of which are built inside private firms. This renders them doubly-inscrutable. As we have already emphasized, we do not understand why or how massive neural networks function, and the mapping from inputs (data and training procedures) to model outputs is mysterious and based largely on trial-and-error. And because frontier models are now trained with considerable secrecy within private labs, we don't even know what goes into such models. Nor are smaller open-weight models good guides---they simply do not display the capabilities and behaviors of larger models \citep{kaplan2020scaling}.

This presents a serious challenge to our regulatory and legal institutions. At present, regulatory implications do not feature as a significant consideration in the legal design of the boundary of the firm. After all, regulators don't need access to the internal processes of automobile or pharmaceutical manufacturers in order to assess their safety and performance---they can
simply test the final products or draw on public domain science to evaluate them. But the massive AI models developed in commercial labs cannot be replicated and evaluated in government or academic labs: the costs of training are too high and evaluation requires access not only to model outputs but also inputs---their data and training procedures. 

For these reasons, governments and the academic researchers that can contribute public domain knowledge to regulatory efforts will need access to information that is now considered proprietary to the firm. Regulators in other domains, of course, routinely gain access to confidential information: pharmaceutical firms have to allow inspectors access to their production facilities to ensure compliance with manufacturing requirements; tax authorities can demand access to a firm's financial records; detailed commercial information can be subpoenaed by antitrust officials in litigation. But in these cases, regulatory authority is based on a policy assessment as to what firms are required to do and hence what information the government has a right to access. In the case of modern AI, however, governments do not know if and how they should regulate. This poses thorny questions economists are well-placed to tackle. We have careful accounts of the economic rationale for patent and copyright law, with attention to the tradeoff between solving the free-rider problem in innovation and the costs of monopoly distortions \citep{gallini_scotchmer_2002}. But we do not yet have a correspondingly robust economic account of trade secret and confidentiality law although some accounts have been offered in the law and economics literature \citep{friedman1991some,chiang2024economic}. 
Of course, legislation is only part of the remedy. Just as  financial firms tend to game stress tests \citep{fed_stress_test_gaming_2016}, AI firms might have considerable leeway to manipulate the information they share, or to flout safety procedures when it conflicts with profit motives. Thus, even if the legal boundaries of firms are made porous, this raises new economic questions about when to inspect and what to look for.\footnote{\cite{varas2020random,ball2023should} study when inspections should be conducted to ensure compliance. \cite{leisten2023dynamic} offers evidence on the efficacy of linked inspections. Computer scientists have recently advocated for `evidence-seeking' AI policy that actively produces new information \citep{casper2025pitfalls} the private sector would not otherwise produce; recent work in dynamic information acquisition could shed light on these questions \citep{sannikov2024exploration}.} 

\section{Concluding remarks}

Silicon Valley promises us increasingly agentic AI systems that might one day supplant human decisions. If this vision materializes, it will reshape markets and organizations with profound consequences for the structure of economic life. But, as we have emphasized throughout this chapter, where we end up within this vast space of possibility is a design choice: we have the opportunity to develop mechanisms, infrastructure, and institutions to shape the kinds of AI agents that are built, and how they interact with each other and with humans. These are fundamentally economic questions---we hope economists will help answer them.  

\footnotesize  
\setstretch{0.8}
\setlength{\bibsep}{-1pt}
\bibliography{ref}

\end{document}